\documentclass[prd,aps,preprintnumbers,showpacs,twocolumn,nofootinbib]{revtex4-1}

\usepackage{amsmath}
\usepackage{amssymb}
\usepackage{graphicx}
\usepackage{epsfig}
\usepackage{color}
\usepackage{url}
\usepackage{times}
\usepackage{bm}
\usepackage{mathrsfs}
\usepackage[utf8]{inputenc}
\usepackage{hyperref}

\newcommand{\beq}{\begin{equation}}
\newcommand{\eeq}{\end{equation}}
\newcommand{\bea}{\begin{eqnarray}}
\newcommand{\eea}{\end{eqnarray}}
\newcommand{\bit}{\begin{itemize}}
\newcommand{\eit}{\end{itemize}}

\newcommand{\dd}{\rm d}

\newcommand{\nn}{\nonumber}

\newcommand{\TPI}{\address{Theoretisch-Physikalisches Institut,
    Friedrich-Schiller-Universit\"at Jena,\\ Max-Wien-Platz 1,
          D-07743 Jena, Germany}}

\begin{document}

\title{Gyromagnetic factor of rotating disks of electrically charged dust in general relativity}
\author{Yu-Chun Pynn}
\author{Rodrigo Panosso Macedo}
\email{rodrigo.panosso-macedo@uni-jena.de}
\author{Martin Breithaupt}
\author{Stefan Palenta}
\author{Reinhard Meinel}

\TPI
\date{\today}

\begin{abstract}
We calculated the dimensionless gyromagnetic ratio (``$g$-factor'') of self-gravitating, uniformly rotating disks of 
dust with a constant specific charge $\epsilon$. These disk solutions to the Einstein-Maxwell equations depend on $\epsilon$ and a ``relativity parameter'' $\gamma$ ($0<\gamma\le 1$) up to a scaling parameter. Accordingly, the 
$g$-factor is a function $g=g(\gamma,\epsilon)$. The Newtonian limit is characterized by $\gamma \ll 1$, whereas 
$\gamma\to 1$ leads to a black-hole limit. The $g$-factor, for all $\epsilon$, approaches the values $g=1$ as 
$\gamma\to 0$ and $g=2$ as $\gamma\to 1$.
\end{abstract}


\maketitle

\section{Introduction}
To any physical system with a well defined notion for the observables mass $M$, angular momentum $J$, electric charge $Q$ and magnetic dipole moment $\mu_{\rm B}$, it is common to introduce the gyromagnetic ratio ($g$-factor)
\beq
\label{eq:g_factor}
{ g}  = 2\dfrac{M}{Q}\dfrac{\mu_{\rm B}}{J}.
\eeq
Such a dimensionless quantity plays an important role in physics. Since this simple measurement is available in both classical and quantum regimes, it allows one to establish connections between several physical theories.

In fact, the $g$-factor was originally introduced in classical electrodynamics~\cite{Landau-Lifshitz2}. Interestingly, for all classical convective systems (where the ratio of charge and mass density is constant, and where the mass and charge elements have equal velocities, which satisfy $v\ll c$), one obtains the value $g  = 1$.  In quantum mechanics though, a different $g$-factor is necessary for explaining experimental results from Zeeman spectroscopy. In the non-relativistic Pauli equation, the value $g  = 2$ for the magnetic moment associated with the electron's spin must be imposed {\it ad hoc}, while it follows automatically from the Dirac equation, i.e., when relativistic effects are included.

The particular value $g = 2$ is found in general relativity as well. The most notable example is probably the Kerr-Newman solution, describing a charged and rotating black hole~\cite{Carter1968}. Later, the authors of \cite{Reina1975} generalized this property and showed that {\em any} electro-vacuum solution to Einstein-Maxwell's equation obtained by an $SU(2,1)$ invariance transformation~\cite{Ernst1968,Harrison1968,Neugebauer:1969wr} from a pure vacuum solution also has the value $g = 2$. The coincidence around the preferred values $g = 2$ usually motivates one to look for a deeper common root between quantum theory and general relativity (see \cite{Pfister:775342} for a recent review).  

More recently, this topic has been further addressed in several physical scenarios. Of particular interest were ``intermediate" objects in general relativity, for which the gravitational fields were weaker than for the black-hole solution, but with non-negligible strong field effects. Pfister and King considered the case of a rotating charged mass shell~\cite{Pfister2002}. Apart from generalising previous studies on this matter~\cite{Cohen1973, Briggs1981,Mustafa1987}, they noticed that $g \approx 2$ is extremely robust, in the sense that this value is obtained in a big part of the mass shell's parameter space.

However, a different result was obtained in~\cite{Novak2003}. After constructing numerical equilibrium configurations of rotating neutron stars, the authors always found the value $g<2$ within the models considered. In particular, the authors observed values around $g\approx 1$ in the Newtonian regime of the solution, while the highest value measured by them was $g\approx 1.9$. A discrepancy to the preferred value $g=2$ is also found in generalized gravity theories~\cite{HOSOYA198444,GIBBONS1986201,Horne1992,RUSSO1995611,DUFF1997161,Ortaggio2006,Aliev2007}, typically due to the absence of a comparable no-hair theorem and the presence of additional fields contributing to the angular momentum of the system.

Electrically charged rotating disks provide us with an interesting scenario to enrich the discussion on this matter. In fact, without taking gravitational effects into account, relativistically rotating disks were dicussed in~\cite{Lynden-Bell2004}. Even though Einsteins's equations are not considered in his framework, the author shows that the electromagnetic fields share some similarities with the ones resulting from the Kerr-Newman solution in the limit of vanishing gravitational constant $G$.  

In this work, we consider the complete self-gravitating set-up in general relativity and we show that the gyromagnetic ratio of rotating disks of electrically charged dust interpolates smoothly between the classical value $g = 1$ up until the black-hole value $g =2$. Note that this system cannot be obtained directly from the known solution of rotating disks of dust~\cite{Bardeen_Wagoner_1971,Neugebauer1995}. In fact, by performing a Harrison transformation~\cite{Ernst1968,Harrison1968,Neugebauer:1969wr}  on the rotating disk solution, one always obtains new (charged) solutions to Einstein-Maxwell's equations with $g = 2$. Yet, the energy-momentum tensor of those new solutions is, in general, not a physically acceptable source~\cite{Klein2002}.   

The construction of our solution follows the strategy from~\cite{Meinel:2012wm,palenta_meinel_2013,Breithaupt:2015xva}. Assuming stationarity and axial symmetry, it consists of solving Einstein-Maxwell's equations for a system with an energy-momentum tensor whose contributions come from the dust particles and from the electromagnetic fields. The system is parametrized in terms of a constant specific charge $\epsilon \in [-1,1]$ and a ``relativity parameter'' $\gamma \in (0,1]$. Based on the algorithm introduced in~\cite{Petroff2001}, the authors of ~\cite{palenta_meinel_2013,Breithaupt:2015xva} were able to calculate the solution in terms of a high order post-Newtonian expansion in the parameter $\gamma$. In particular,~\cite{Breithaupt:2015xva} provided strong evidence that, analogous to the uncharged case~\cite{Kleinwachter:2010di}, the limit $\gamma\rightarrow 1$ leads to the extreme Kerr-Newman black hole.

Contrary to the post-Newtonian expansion from~\cite{palenta_meinel_2013,Breithaupt:2015xva}, we here resort to numerical methods in order to obtain a (highly) accurate solution around the black-hole limit $\gamma\rightarrow 1$. To this end, we make use of a (pseudo-)spectral method, whose algorithm is based on the one described in~\cite{Meinel:2008}. 

This paper has the following structure: section \ref{sec:PhysModel} introduces the physical model. It discusses the field equations and the parameter space of the system. Section \ref{sec:NumMeth} is devoted to the numerical method employed in this work. Section \ref{sec:Results} then presents our results, while section \ref{sec:Discussion} summarizes this work and brings some future perspectives. We use the following conventions: boldface letters denote the abstract representation of tensors while latin indices $a,b,\cdots$ are used to express their components in a given coordinate basis $\{ \partial_a\}$. Moreover, latin indices in parentheses $(a),(b),\cdots$ refer to the components of a tensor in a given tetrad basis $\mathbf{e}_{(a)} = e_{(a)}{}^a\partial_a$. We use units in which $G=c=4\pi\epsilon_0=1$.    

\section{Rotating disk of charged dust}\label{sec:PhysModel}

\subsection{Geometrical setup}
The charged disk is completely described by Einstein's field equations
\beq
\label{eq:EinsteinEq}
R_{ab} = 8\pi \left[ T_{ab} - \dfrac{1}{2}g_{ab} T \right] \quad {\rm with} \quad T = T_a{}^a,
\eeq
together with Maxwell's equations
\beq
\label{eq:MaxwellEq}
\nabla_b F^{ab} = 4 \pi j^a, \quad \dd {\mathbf F} = 0.
\eeq
With the assumption of stationarity and axial symmetry through the existence of Killing vectors $\boldsymbol{\xi}$ and $\boldsymbol{\eta}$, we can globally express the metric in terms of the Weyl-Lewis-Papatetrou coordinates $\{ t, \rho, \zeta, \phi \}$ as  
\beq
\label{eq:Metric}
\dd s^2 = \alpha^2 \left[ \dd \rho^2 + \dd \zeta^2 \right] + \dfrac{\rho^2}{\nu^2} \left[ \dd \phi - \omega\, \dd t \right]^2 - \nu^2 dt^2,
\eeq
where the unknown functions $\alpha, \nu$ and $\omega$ depend only on the coordinates $\{ \rho, \zeta \}$. In this adapted coordinated system, the Killing vectors assume the simple form $\boldsymbol{\xi} = \partial_t$ and $\boldsymbol{\eta} = \partial_\phi$. We remark that the line element \eqref{eq:Metric} has a slightly different representation than the one used in~\cite{Meinel:2012wm,palenta_meinel_2013,Breithaupt:2015xva}.

Besides, the homogenous Maxwell equation in \eqref{eq:MaxwellEq} is trivially satisfied with the introduction of the vector potential $A_a$ via
\bea
F_{ab} = \nabla_a A_b - \nabla_b A_a. \nn 
\eea
The vector potential can be put in the form
\bea
{\mathbf A} = A_t(\rho, \zeta) \, \dd t + A_{\phi}(\rho, \zeta) \, \dd \phi 
\eea
due to the axial symmetry.

Finally, it will be useful to introduce a tetrad basis $\{ \mathbf{e}_{(a)} \}$ as proposed in \cite{Bardeen_Wagoner_1971}
\bea
\label{eq:ZAMOtetrad}
\mathbf{e}_{(0)} &=& \dfrac{1}{\nu}\left[ \partial_t + \omega \, \partial_\phi \right], 
\quad \mathbf{e}_{(1)} = \dfrac{1}{\alpha}\, \partial_\rho, \nn \\
\quad \mathbf{e}_{(2)} &=& \dfrac{1}{\alpha}\, \partial_\zeta, \quad \mathbf{e}_{(3)} = \dfrac{\nu}{\rho}\, \partial_\phi.
\eea
Since $e_{(0)}{}^a\eta_a = 0$, this tetrad is related to the local inertial frame of zero angular momentum observers.

\subsection{Model of matter}
The energy-momentum tensor $T_{ab}$ is composed by a dust and an electromagnetic (EM) contribution, i.e., $T_{ab} = T_{ab}^{\rm dust} + T_{ab}^{\rm EM}$, with
\bea
T_{ab}^{\rm dust} &=& \mu u_{a}u_{b} \quad {\rm and} \nn \\
\label{eq:EnergyMomentTensor}
 T_{ab}^{\rm EM} &=& \dfrac{1}{4\pi}\left( F_{ac}F_b{}^c - \dfrac{1}{4} g_{ab} F_{cd}F^{cd}   \right).
\eea
In the expressions above, $\mu$ is associated to the baryonic mass density of the dust particles, while $u^a$ describes their $4-$velocity. In the coordinate system $\{t, \rho, \zeta, \phi \}$, we consider the disk at the equatorial plane $\zeta = 0$, with a range $\rho\in [0,\rho_0]$ and therefore the baryonic mass density assumes the form
\beq
\mu = \dfrac{ \sigma_{\rm P}}{\alpha}\delta(\zeta),
\eeq 
with $\delta(\zeta)$ the Dirac delta and $\sigma_{\rm P}(\rho)$ the proper surface mass density~\cite{Meinel:2008}. The disk's coordinate radius $\rho_0$ sets the length scale of the system.

The $4-$velocity is expressed in terms of the Killing vectors $\boldsymbol{\xi}$ and $\boldsymbol{\eta}$ as
\beq
u^a = \dfrac{1}{\nu \sqrt{1-V^2}}\left[ \delta^a_t + \Omega \, \delta^a_\phi \right], \quad {\rm with} \quad V = \dfrac{\rho}{\nu^2}\,(\Omega - \omega).
\eeq
Here, $\displaystyle \Omega = \dfrac{\dd \phi}{\dd t}$ is the dust particle angular velocity. The quantity $V$ ensures the normalisation $u^a u_a = -1$ and it can be physically interpreted as the relative velocity between the dust particle and a zero angular momentum observer\footnote{In fact, in terms of the tetrad basis \eqref{eq:ZAMOtetrad}, $\mathbf{u}$ results from the boost  $\displaystyle \mathbf{u} = \dfrac{1}{\sqrt{1-V^2}}\left[ \mathbf{e}_{(0)} + V\,\mathbf{e}_{(3)} \right]$. }. In this work, we are interested in disks with rigid rotation, i.e., with $\Omega$ constant.

For the charged particles, we assume a purely convective $4-$current density
\beq
j^a = \varrho_{\rm el} u^a \quad {\rm with} \quad \varrho_{\rm el} = \epsilon \mu,
\eeq
i.e.,~the charge density $\varrho_{\rm el}$ is related to the mass density via the constant specific charge $\epsilon\ \in [-1,1]$. 

\subsection{Field equations and boundary conditions}\label{sec:Eq_BC}
The field equations are conveniently expressed in terms of the tetrad basis ${\mathbf e}_{(a)}$. Let 
\bea
{\cal E}_{(a)(b)} &=& e_{(a)}{}^ae_{(b)}{}^b\left [R_{ab} - 8\pi \left( T_{ab} - \dfrac{1}{2}g_{ab} T \right) \right] \\
{\cal M}_{(a)} &=& e_{(a)}{}^a\left [\nabla_b F_{a}{}^{b} - 4 \pi j_a \right] 
\eea
be the projection of Einstein's equations \eqref{eq:EinsteinEq} and Maxwell equations \eqref{eq:MaxwellEq} into the basis \eqref{eq:ZAMOtetrad}. Then, we obtain from the components ${\cal E}_{(0)(0)}, {\cal E}_{(0)(3)}, {\cal M}_{(0)}$ and ${\cal M}_{(3)}$ 
\begin{widetext}
\bea
\label{eq:EOM1}
 &\Delta \nu - \dfrac{|\nabla \nu|^2}{\nu} - \dfrac{1}{2}\dfrac{\rho^2}{\nu^3}|\nabla \omega|^2 - \dfrac{1}{\nu}\left[ \omega\nabla A_{\phi} + \nabla  A_{t} \right]^2 - \dfrac{\nu^3}{\rho^2}|\nabla A_{\phi}|^2 = 4\pi\sigma_{\rm P}\alpha\, \nu \,\dfrac{1+V^2}{1-V^2}, \\
\label{eq:EOM2}
& \nabla  \left[ \dfrac{\rho^2}{\nu^4} \nabla\omega \right] - \dfrac{4}{\nu^2}\nabla A_\phi \cdot \left[ \omega \nabla A_\phi + \nabla A_t \right] = -16\pi\sigma_{\rm P}\alpha \, \dfrac{\rho}{\nu^2}\, \dfrac{V}{1-V^2}, \\
\label{eq:EOM3}
& \nabla\left[ \dfrac{\nu^2}{\rho^2} \nabla A_\phi \right] - \dfrac{\nabla \omega}{\nu^2}\cdot\left[ \omega \nabla A_\phi + \nabla A_t  \right] = -4\pi\sigma_{\rm P}\alpha \, \epsilon \, \dfrac{\nu}{\rho} \, \dfrac{V}{\sqrt{1-V^2}}, \\
\label{eq:EOM4}
&\nabla\left[ \dfrac{1}{\nu^2} \left( \omega \nabla A_\phi + \nabla A_t \right)\right] = \dfrac{ 4\pi\sigma_{\rm P}\alpha \, \epsilon}{\nu} \, \dfrac{1}{\sqrt{1-V^2}}.
\eea
\end{widetext}
The symbols $\nabla$ and  $\Delta$ respectively, denote the usual gradient and Laplacian operators in flat space, expressed here in cylindrical coordinates $\{\rho, \zeta, \phi \}$. 

Outside the disk range (in the electro-vacuum region), we have $\sigma_{\rm P} = \epsilon = 0$ and the right-hand sides of equations \eqref{eq:EOM1}-\eqref{eq:EOM4} vanish. Hence, we obtain a coupled system of four elliptic equations for the four variables $\nu, \omega, A_t$ and $A_\phi$. Once these fields are known, one can use the remaining equations ${\cal E}_{(1)(1)}$ and ${\cal E}_{(1)(2)}$ to obtain $\alpha$. 

In order to uniquely solve the system of elliptic equations  \eqref{eq:EOM1}-\eqref{eq:EOM4}, we need to specify boundary conditions that describe the physical scenario we want to model. Concretely, there are four surfaces of interest (see figure \ref{fig_boundary_conditions_in_rho_zeta}):

\bit
\item Region $\cal A$: the symmetry axis $(\rho = 0, \zeta \neq 0)$
\eit
Equations \eqref{eq:EOM1}-\eqref{eq:EOM4} impose the following regularity conditions
\bea
\label{eq:BoundA}
\nu_{,\rho}(0, \zeta) &=& 0,  \quad
\omega_{,\rho}(0, \zeta) = 0, \\
A_t{}_{,\rho}(0, \zeta) &=& 0, \quad
A_\phi{}(0, \zeta) = 0. \nn
\eea

\bit
\item Region $\cal B$: spacelike infinity $(r = \sqrt{\rho^2 + \zeta^2 } \rightarrow \infty)$
\eit
We demand the physical condition of asymptotic flatness
\bea
\label{eq:BoundB}
\lim_{r\rightarrow \infty}\nu(\rho, \zeta) &=& 1, \quad
\lim_{r\rightarrow \infty}\omega(\rho, \zeta) = 0, \\
\lim_{r\rightarrow \infty}A_t(\rho, \zeta) &=& 0, \quad
\lim_{r\rightarrow \infty}A_\phi(\rho, \zeta) = 0. \nn
\eea

\bit
\item Region $\cal C$: equatorial plane without matter $(\rho>\rho_0, \zeta = 0)$
\eit
Equatorial symmetry imposes

\bea
\label{eq:BoundC}
\nu_{,\zeta}(\rho, 0) &=& 0, \quad 
\omega_{,\zeta}(\rho, 0) = 0,  \\
A_t{}_{,\zeta}(\rho, 0) &=& 0, \quad
A_\phi{}_{,\zeta}(\rho, 0) = 0. \nn
\eea

\bit
\item Region $\cal D$: disk of charged dust $(\rho\in[0,\rho_0], \zeta = 0)$
\eit

The surface mass density $\sigma_{\rm P}$ introduces a discontinuity in the first derivative along $\zeta$. Integrating equations \eqref{eq:EOM1}-\eqref{eq:EOM4} along $\zeta\in [-z , + z]$ with $z\rightarrow 0$, we obtain

\bea
\label{eq:BoundD}
& \nu_{,\zeta}(\rho, 0^+) = 2\pi\sigma_{\rm P}\alpha\,\nu \, \dfrac{1+V^2}{1-V^2}, \nn \\ 
& \omega_{,\zeta}(\rho, 0^+) = -8\pi\sigma_{\rm P}\alpha \, \dfrac{\nu^2}{\rho}\, \dfrac{V}{1-V^2} \\
& {A_{\phi}}_{,\zeta}(\rho, 0^+) = -2\pi\sigma_{\rm P}\alpha \, \epsilon \, \dfrac{\rho}{ \nu} \, \dfrac{V}{\sqrt{1-V^2}}, \nn \\ 
& {A_t}_{,\zeta}(\rho, 0^+) + \omega {A_\phi}_{,\zeta}(\rho, 0^+) = 2\pi\sigma_{\rm P}\alpha \, \epsilon \, \nu\, \dfrac{1}{\sqrt{1-V^2}}.
\nn
\eea
Since $\alpha$ is decoupled from the other fields, we can eliminate this quantity from the boundary conditions by combing any two of the equations in \eqref{eq:BoundD}, which yields
\bea
\label{eq:BoundD2}
& \nu_{,\zeta}(\rho, 0^+) = - \dfrac{\rho }{4 \nu} \, \dfrac{1+V^2}{V}\, \omega_{,\zeta}(\rho, 0^+), \nn \\
& {A_\phi}_{,\zeta}(\rho, 0^+) =\epsilon \,  \dfrac{\rho^2}{4 \nu^3} \,   {\sqrt{1-V^2}}\, \omega_{,\zeta}(\rho, 0^+), \quad  \\
& {A_t}_{,\zeta}(\rho, 0^+) + \omega{A_\phi}_{,\zeta}(\rho, 0^+) = - \epsilon  \dfrac{\rho}{4 \nu} \dfrac{\sqrt{1-V^2}}{V} \,\omega_{,\zeta}(\rho, 0^+).
\nn
\eea
The boundary conditions \eqref{eq:BoundD2} are complemented with a relation following from $\nabla_b T^{ab} = 0.$ The divergence-free condition of the energy-momentum tensor is easily interpreted if one considers the $4-$velocity $u^a$ and its associated projection operator $h_{ab} = g_{ab} +u_a\,u_b$. In fact, the contraction $u_a\nabla_b T^{ab}$ leads to the conservation of the baryonic mass $\nabla_a(\mu u^a) = 0$, while $h_{ab}\nabla_c T^{bc}$ gives
\beq
\label{eq:NewtonLaw}
f_a = \mu a_a,
\eeq
with the acceleration $a_a = u^b\nabla_b u_a$ and the Lorentz force $f_a = h_{ab}F^{bc}j_c$. The $\rho-$component of \eqref{eq:NewtonLaw} reads\footnote{This condition also follows from a convenient combination of the equations ${\cal E}_{(1)(1)}, {\cal E}_{(1)(2)}$ and ${\cal E}_{(a)}{}^{(a)}$.}
\beq
\label{eq:BoundD3}
(1+V^2)\nu_{,\rho} = \left[ \dfrac{V}{\rho} \nu - \dfrac{\rho}{\nu}\omega_{,\rho}\right]V
+  {\epsilon}\sqrt{1-V^2}\left[ A_t{}_{,\rho} + \Omega A_\phi{}_{,\rho} \right].
\eeq
As discussed in~\cite{Meinel:2012wm,palenta_meinel_2013,Breithaupt:2015xva},  one can integrate equation \eqref{eq:BoundD3} in cases where $\Omega$ and $\epsilon$ are constant to obtain
\beq
\label{eq:Parameter_Relation}
D := \nu\sqrt{1-V^2} - \epsilon \left( A_t + \Omega A_\phi \right) = {\rm constant.}
\eeq
The value of the constant $D$ is obtained by inspecting the right-hand side of eq.~\eqref{eq:Parameter_Relation} at any value $\rho \in [0,\rho_0]$ and $\zeta = 0$. Concretely, at the center of the disk ($\rho=0$), we obtain
\beq
\label{eq:ConstantD}
D = \nu^{\rm c} -  \epsilon A_t^{\rm c},
\eeq 
with $\nu^{\rm c} = \nu(\rho = 0, \zeta =0)$ and $A_t^{\rm c} = A_t(\rho = 0, \zeta =0)$.

Note that the boundary condition in the differential form \eqref{eq:BoundD3} provides us with a more generic set-up than the version in eq.~\eqref{eq:Parameter_Relation}. In fact, \eqref{eq:BoundD3} could also be used to model disks with a differential rotation $\Omega = \Omega(\rho)$, whereas \eqref{eq:Parameter_Relation} is restricted to the rigid rotation case $\Omega =$constant.

Finally, let us remark that eq.~\eqref{eq:BoundD3} fixes the field $\nu$ at the disk up to the integration constant $D$. In order to solve the equations numerically, it is crucial to assert that the system has a unique solution. Therefore, at the point $(\rho=0, \zeta=0)$ one would have to fix the value of the integration constant. Equivalently (and more convenient from the physical point of view, see discussion in the next section), one can specify a given value for the quantity $\nu^{\rm c}.$ 
  
\subsection{Parameter space and physical quantities}
The parameter space of the problem has been identified in the works~\cite{Meinel:2012wm,palenta_meinel_2013,Breithaupt:2015xva}. In our system of units, the specific charge assumes values in the range $-1\leq \epsilon \leq 1$. Two values of this parameter are of particular relevance. The solution to the (uncharged) disk of dust~\cite{Bardeen_Wagoner_1971,Neugebauer1995,Petroff2001} is clearly recovered in the case $\epsilon = 0$. On the other hand, the case $|\epsilon| = 1$ leads to the so-called electrically counterpoised dust configuration (see, e.g., \cite{Meinel:2011ur}), in which the gravitational attraction is exactly counter-balanced by the electric repulsion. 

Apart from the specific charge $\epsilon$ (without loss of generality, we restrict ourselves to $\epsilon \geq 0$), it is convenient to introduce the relativity parameter 
\beq 
\label{eq:gamma}
\gamma = 1 - \nu^{\rm c},
\eeq 
also used in the study of the uncharged disk~\cite{Bardeen_Wagoner_1971,Neugebauer1995,Petroff2001,Kleinwachter:2010di}. This parameter is related to the redshift $Z_{\rm c}$ of a photon emitted at the centre of the disk and measured at infinity via $\gamma = Z_{\rm c}/(1+Z_{\rm c})$. As in the uncharged case, one intuitively expects to obtain the Newtonian limit as $\gamma \ll 0$, while $\gamma \rightarrow 1$ should lead to a black-hole transition. Indeed, first studies of the post-Newtonian expansion provide a strong indication for this behavior~\cite{Breithaupt:2015xva}. 

With such a parametrisation, the angular velocity $\Omega$ is not a free quantity that we are allowed to choose. Since $\Omega$ depends on the freely specifiable parameters $\{\gamma,\epsilon\}$, it must be considered as an unknown variable. Therefore, the numerical scheme should be able to account for this extra unknown parameter together with the field variables (see discussion in section \ref{sec:NumMeth}). Apart from $\Omega$, we are interested in the dependence of the following physical quantities upon the parameters $\{\gamma, \epsilon\}$: the mass $M$, angular moment $J$, electric charge $Q$ and magnetic moment $\mu_{\rm B}$. In terms of a spherical-type representation of the coordinates $\rho = r \sin\theta$ and $\zeta = r\cos{\theta}$, these observables are computed out of the far-field behavior of the field variables via
\beq
\label{eq:FarFieldParameters}
\nu \sim 1 -\dfrac{M}{r}, \, 
\omega \sim \dfrac{2J}{r^3}, \, 
A_t \sim -\dfrac{Q}{r} \quad {\rm and} \quad
A_\phi \sim \dfrac{\mu_{\rm B}}{r}\sin^2\theta.
\eeq
The gyromagnetic factor $g$ is then directly obtained according to \eqref{eq:g_factor}.
The physical quantities derived from the far-field are connected to the disk quantities by the relation \cite{Meinel:2012wm,palenta_meinel_2013,Breithaupt:2015xva}
\bea
M&=& 2\Omega J+ D \frac{Q}{\epsilon} \nn \\
\label{eq:FarFieldNearField}
   &=& 2 \Omega J + \left[\frac{1-\gamma}{\epsilon}-A_t^{\rm c}\right]{Q},
\eea
with the second line obtained from~\eqref{eq:ConstantD} and \eqref{eq:gamma}. Since eqs.~\eqref{eq:FarFieldParameters} and \eqref{eq:FarFieldNearField} are derived independently from each other, the latter provides us with a solid test for the correctness of our framework. 

\section{Numerical Methods}\label{sec:NumMeth}
\subsection{Adapted coordinates}
In order to use spectral methods to solve the set of equations \eqref{eq:EOM1}-\eqref{eq:EOM4}, we first need to map the original domain $[\rho, \zeta]\in [0,\infty)\times (-\infty\times \infty)$ into a compact region $(\sigma, \tau) \in [0,1]^2$. The aim is that the regions ${\cal A}, {\cal B}, {\cal C}$ and ${\cal D}$ are mapped into the boundaries of the numerical domain. This objective is achieved by two coordinate transformations
\begin{figure}[t]
\begin{center}
\includegraphics[width=8.0cm]{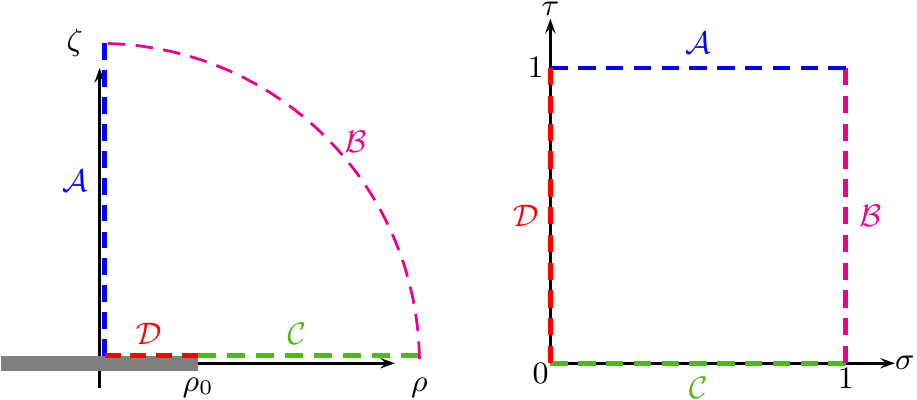}
\end{center}
\caption{The rotating disk of charged dust shown in the Weyl coordinates (left) and the compactified coordinates (right). The thick line denotes the infinitely thin disk with radius $\rho_0$. The areas illustrated in the figure imply each particular part of the boundary conditions: $\mathcal{A}$: $\zeta$-axis, $\mathcal{B}$: Infinity, $\mathcal{C}$: Equatorial plane outside of the disk, $\mathcal{D}$: On the disk surface. }
\label{fig_boundary_conditions_in_rho_zeta}
\end{figure}
\bea
\rho &=& \rho_0\sqrt{1+\xi^2}\sqrt{1-\eta^2}, \quad \zeta = \rho_0\xi\eta \quad {\rm and} \nn \\ 
\label{eq:SpecCoord}
\sigma &=& \dfrac{2}{\pi}\arctan{\xi}, \quad \tau = \eta^2.
\eea
The former introduces the elliptic coordinates $(\xi, \eta)\in [0, \infty]\times[0,1]$, while the latter compactifies the $\xi-$direction. Note that we exploit the equatorial symmetry and restrict ourselves to the region $\zeta  \ge 0$ ($\eta \ge 0 $). Altogether,  we obtain the following maps (see figure \ref{fig_boundary_conditions_in_rho_zeta}): 
\bit
\item Region ${\cal A}: \eta =  1 \Rightarrow \tau = 1$ 
\item Region ${\cal B}: \xi \rightarrow \infty \Rightarrow \sigma = 1$
\item Region ${\cal C}: \eta = 0 \Rightarrow \tau = 0$
\item Region ${\cal D}: \xi = 0 \Rightarrow \sigma = 0$.
\eit
In appendix \ref{app:Eqs_SpecCoord}, we explicitly give the corresponding expression for the field equations \eqref{eq:EOM1}-\eqref{eq:EOM4} and the boundary conditions  \eqref{eq:BoundA}-\eqref{eq:BoundC}, \eqref{eq:BoundD2} and \eqref{eq:BoundD3} in terms of the spectral coordinates $\{\sigma, \tau \}$.

\subsection{Spectral Methods}
\label{Spectral_Methods}
As already mentioned, we solve the field equations by means of a (pseudo-)spectral method and here we give some details on the techniques used. Let us recall that, apart from the functions $\nu(\sigma, \tau), \omega(\sigma, \tau), A_t(\sigma, \tau)$ and $A_\phi(\sigma, \tau)$, we also must include the parameter $\Omega$ as an unknown in our scheme. As usual in any spectral algorithm, we first fix a resolution $N_{\sigma}$ and $N_{\tau}$ and consider a vector $\vec{X}$ composed of all the variables of the system
\beq
\vec{X} = \left( \nu^{ij} \, \omega^{ij}\, A_t^{ij}\, A_\phi^{ij}\, |\, \Omega \right)^{\rm T} \, {\rm for} \,\, {i = 0\cdots N_{\sigma}, j = 0\cdots N_{\tau}}.
\eeq
In the above expression, we use the notation\footnote{With $f$ denoting either $\nu, \omega, A_t$ or $A_\phi$}  $f^{ij} = f(\sigma_i, \tau_j)$ to denote the function values at the Chebyschev-Lobatto grid points given by
\beq
\label{eq:GridPoints}
\sigma_i = \dfrac{1}{2}\left[ 1 + \cos\left( \pi\dfrac{i}{N_\sigma}\right)\right], \,
\tau_j = \dfrac{1}{2}\left[ 1 + \cos\left( \pi\dfrac{j}{N_\tau}\right)\right].
\eeq
For each function $f$ stored in $\vec{X}$, we can compute its corresponding Chebyshev coefficients $c^{mn}$ by inverting the relation
\beq
f^{ij} = \sum_{m=0}^{N_\sigma}\sum_{n=0}^{N_\tau} c^{mn}\, T_m(2\sigma_i-1)\,T_n(2\tau_j-1).
\eeq
Finally, we compute spectral approximations of first and second derivatives in the $\sigma-$ and $\tau-$directions at all grid points \eqref{eq:GridPoints} which we perform by applying specific differentiation matrices to the vector $\vec{X}$, see \cite{Boyd00,canuto_2006_smf}. 

With all the discrete quantities available, we evaluate the field equations \eqref{eq:EOM1_newcoord}-\eqref{eq:EOM4_newcoord} and boundary conditions \eqref{eq:BC_A_newcoord}-\eqref{eq:BC_D_newcoord} at the grid points \eqref{eq:GridPoints}. This set of equations+boundary conditions forms a system for determining the field variables $\nu, \omega, A_t$ and $A_\phi$. We still need one extra condition to fix the parameter $\Omega$ uniquely, which is achieved by explicitly imposing the value of $\nu^{\rm c} = 1 - \gamma$ at the center of the disk [see eq.~\eqref{eq:ExtraCond}]. Altogether, we obtain a non-linear system of algebraic equations $\vec{F}(\vec{X})$ of order $n_{\rm total} = 4(N_\sigma+1)(N_\tau+1) + 1$. This system is solved with a Newton-Raphson scheme. Note that within the Newton-Raphson scheme, one must solve a linear system involving the Jacobian matrix $\hat{J} = \partial{\vec{F}}/\partial{\vec{X}}$. As detailed in \cite{Meinel:2008}, this linear system is solved with the iterative BiCGStab 
method, with a pre-conditioner based on a finite difference representation of the algebraic system of equations. 

In order to cover all the parameter space $\{ \gamma, \epsilon\}$, we start with parameters $\gamma \sim 0$ and $\epsilon=0$ and provide the solver with a initial guess $\vec{X}^0$ constructed out of the lowest post-Newtonian approximation
\bea
\nu^2 &=& {1+ 2 U}, \quad \Omega^2=\gamma\left( 1-\dfrac{\gamma}{2} \right) \nn \\
\omega &=& A_t = A_\phi = 0. \nn
\eea
The potential $U$ corresponds to the exact solution for the gravitational potential of the uncharged disk of dust in the Newtonian theory of gravity
\bea
U &=& -\dfrac{4}{3\pi}\Omega^2\rho_0^2 \Bigg\{ {\rm arccot}\,{\xi} \nn \\
	&+&\dfrac{3}{4}\left[ \xi -\left( \xi^2 + \dfrac{1}{3}\right){\rm arccot}\,{\xi}\right](1-3\eta^2)  \Bigg\}. \nn
\eea
Once a solution is available, we use it as an initial-guess for a modified set of parameters $\{ \gamma, \epsilon\}.$ By slowly increasing $\gamma$ and $\epsilon$ we are able to cover the region $(\gamma, \epsilon) \in (0,1)\times[0,1)$ in the parameter space.

We end this section by mentioning that near the ultra-relativistic limit $\gamma = 1$, the functions develop strong gradients around the boundary $\sigma=1$. In order to avoid a massive increase in the resolution $N_\sigma$ (which in turn signficantly slows down the speed of the solver), we implement the analytical mesh-refinement 
\beq
\label{eq:mesh_refinement}
\sigma = 1 - \dfrac{\sinh[ \kappa(1- \bar{\sigma})]}{\sinh(\kappa)} {\quad} {\rm with} {\quad} \kappa \sim |\ln(1-\gamma)|
\eeq 
introduced in~\cite{Meinel:2008} and successfully applied in many different contexts~\cite{Macedo:2014bfa,Ammon:2016szz}. 

\section{Results}\label{sec:Results}
\subsection{Numerical Accuracy}

 \begin{figure}[t]
\begin{center}
\includegraphics[width=8.0cm]{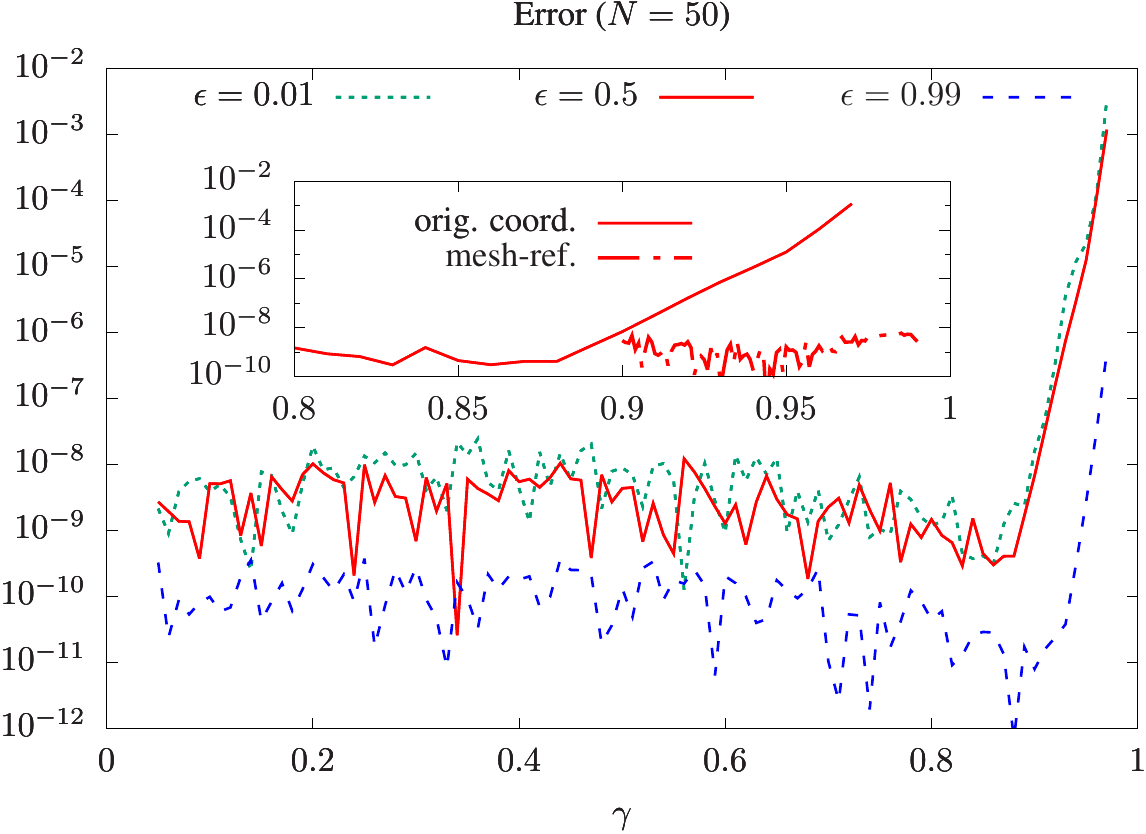}
\end{center}
\caption{ Accuracy test of the physical quantities using \eqref{eq:Phy_Quan_error}. Starting from $\gamma \approx 0.9$, the analytical mesh-refinement is applied to rectify the gradient problem of the field equations around the boundaries. As shown in the inset, high numerical accuracy is assured for values of $\gamma \rightarrow 1$, i.e, in the black-hole limit.}
\label{fig:NumericalError}
\end{figure}

We begin the results section with a technical discussion on the performance of the numerical solution. Note that eq.~\eqref{eq:FarFieldNearField} provides us with a neat accuracy test to check our results. Indeed, the equation relates far field observables (out of which the gyromagnetic factor is constructed) with quantities defined on the disk. Furthermore, it includes the angular velocity $\Omega$, which is an unknown variable within the numerical code on its own. Thus, we introduce an error measurement for the numerical solution via the relative deviation  

 \beq
 \label{eq:Phy_Quan_error}
 {\rm Error} =\left|1-\left[ \frac{2 \Omega J}{M} + \left(\frac{1-\gamma}{\epsilon}-A_t^{\rm c}\right)\frac{Q}{M} \right]\right|.
 \eeq

The error dependence on the parameter $\gamma$ is shown in fig.~\ref{fig:NumericalError} for some representative values of the specific charge $(\epsilon=0.01, 0.5, 0.99)$. The numerical solutions were obtained with a resolution\footnote{We systematically used the same resolution for all parameters discussed. For small values of $\gamma$, however, the numerical saturation could be reached with a smaller number of grid points.} $N_{\sigma} = N_{\tau} = 50$. We observe that the error is of order $\lesssim 10^{-8}$ in a large range of the parameter space. From $\gamma \approx 0.9$ onwards, the error increases significantly due to strong gradients in the fields around the boundary $\sigma=1$. As discussed in section \ref{Spectral_Methods}, we apply the analytical mesh-refinement \eqref{eq:mesh_refinement} to subdue this problem. As shown in the inset of the same figure, this technique is essential to keep the accuracy at $\lesssim10^{-8}$ without a massive increase of the numerical resolution.

The saturation of the numerical resolution at this order of magnitude is limited by the machine precision and it is compatible with the measured observables. Note that the angular momentum is given by $J = \lim\limits_{r\rightarrow \infty} r^3\omega/2$. When expressed in terms of the coordinates $\{\sigma,\tau \}$, the limit can be explicitly performed and it involves third derivatives $\omega_{,\sigma\sigma \sigma}$. The final accuracy is, hence, restricted to the numerical errors on the performance of third derivatives with spectral methods. 

 \subsection{Gyromagnetic factor}
 
\begin{figure}[t]
\begin{center}
\includegraphics[width=8.5cm]{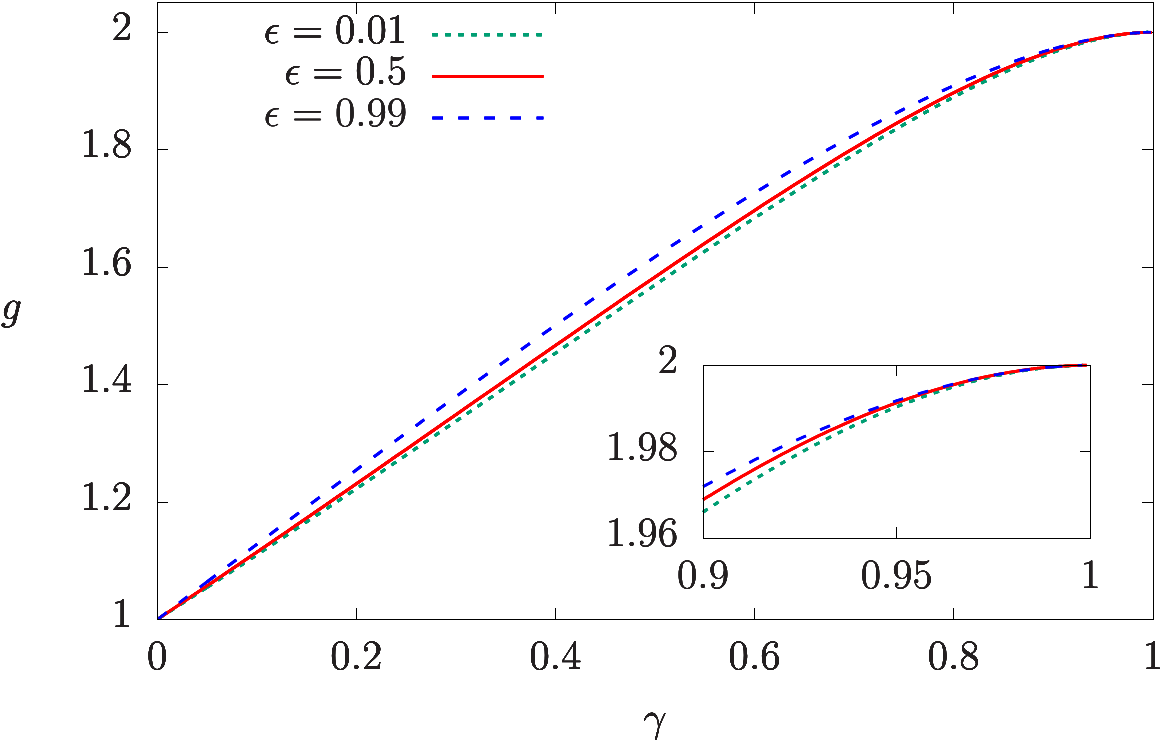}
\end{center}
\caption{Gyromagnetic factor ${ g}$ from the Newtonian to the ultra-relativistic limit with different specific charge $\epsilon$. The lower right window shows the result near the ultrarelativistic limit, with all curves tending monotonically to $g=2$.}
\label{fig:gyro_versus_gamma}
\end{figure}

With the numerical solution under control, we proceed and study the dependence of the gyromagnetic factor on the ``relativity parameter" $\gamma$. Here again, we concentrate ourselves on representative values $\epsilon=0.01, 0.5, 0.99$ for the specific charge. Fig.~\ref{fig:gyro_versus_gamma} confirms that the $g$-factor of rotating disks of electrically charged dust interpolates smoothly between the classical value $g = 1$ and the black-hole limit $g = 2$. Moreover, we obtain a very mild dependence on $\epsilon$. As shown in the figure, the slightly charged case $\epsilon=0.01$ and the near electrically counterpoised case $\epsilon=0.99$ do not deviate drastically from each other.
 
It is interesting to note that the $g$-factor has a (non-vanishing) finite limit in both cases $\epsilon \rightarrow 0$ and $\epsilon \rightarrow 1$. The former corresponds to the uncharged rotating disk with $Q = \mu_{\rm B} = 0$, while the latter leads to the electrically counterpoised case with $J = \mu_{\rm B} = 0$. That the gyromagnetic ratio has a finite limiting case, in spite of vanishing observables, can be best appreciated with the help of the post-Newtonian expressions. In~\cite{Breithaupt:2015xva}, it is shown that the charge, the angular momentum and the magnetic moment scale as $Q \sim \epsilon$, $J\sim \sqrt{1-\epsilon^2}$ and $\mu_{\rm B}\sim\epsilon\sqrt{1-\epsilon^2}$ respectively. Therefore, the ratio $\mu_{\rm B}/(QJ)$ is finite in both limits.

 \begin{figure*}[t]
\begin{center}
\includegraphics[width=8.5cm]{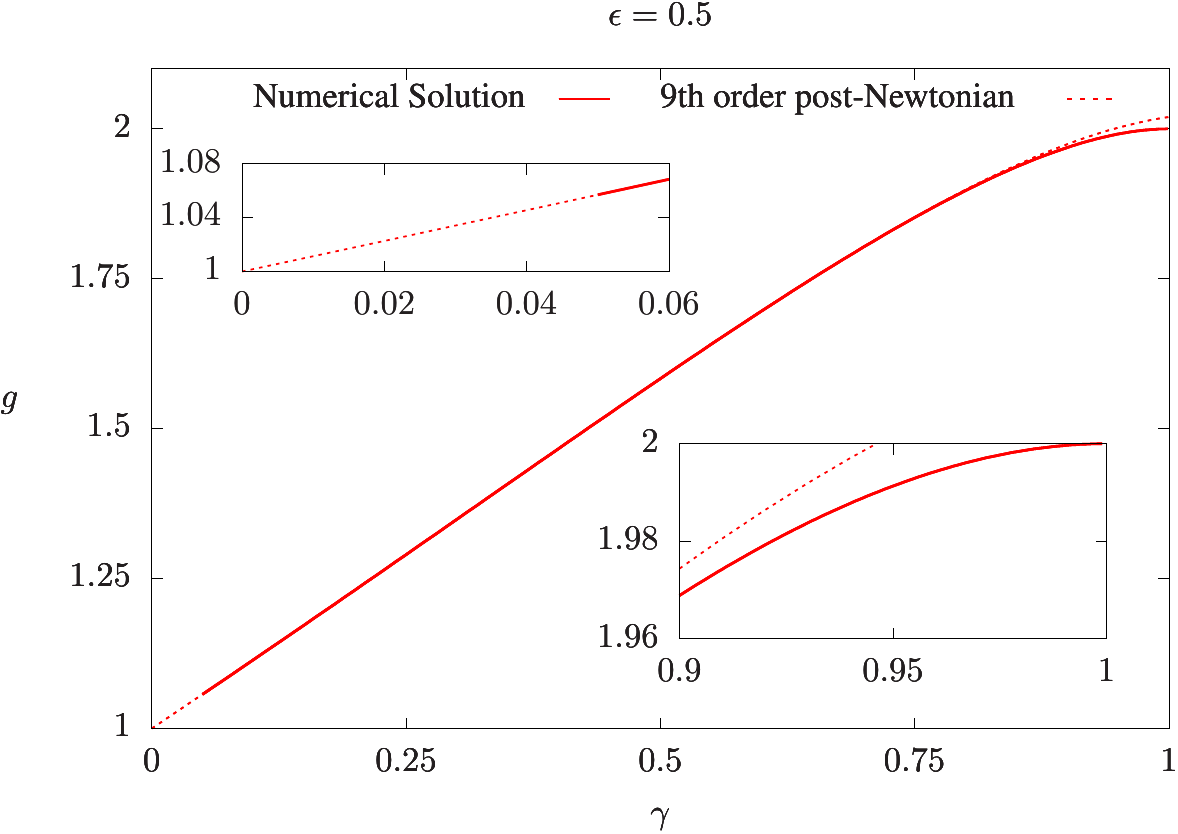}
\includegraphics[width=8.5cm]{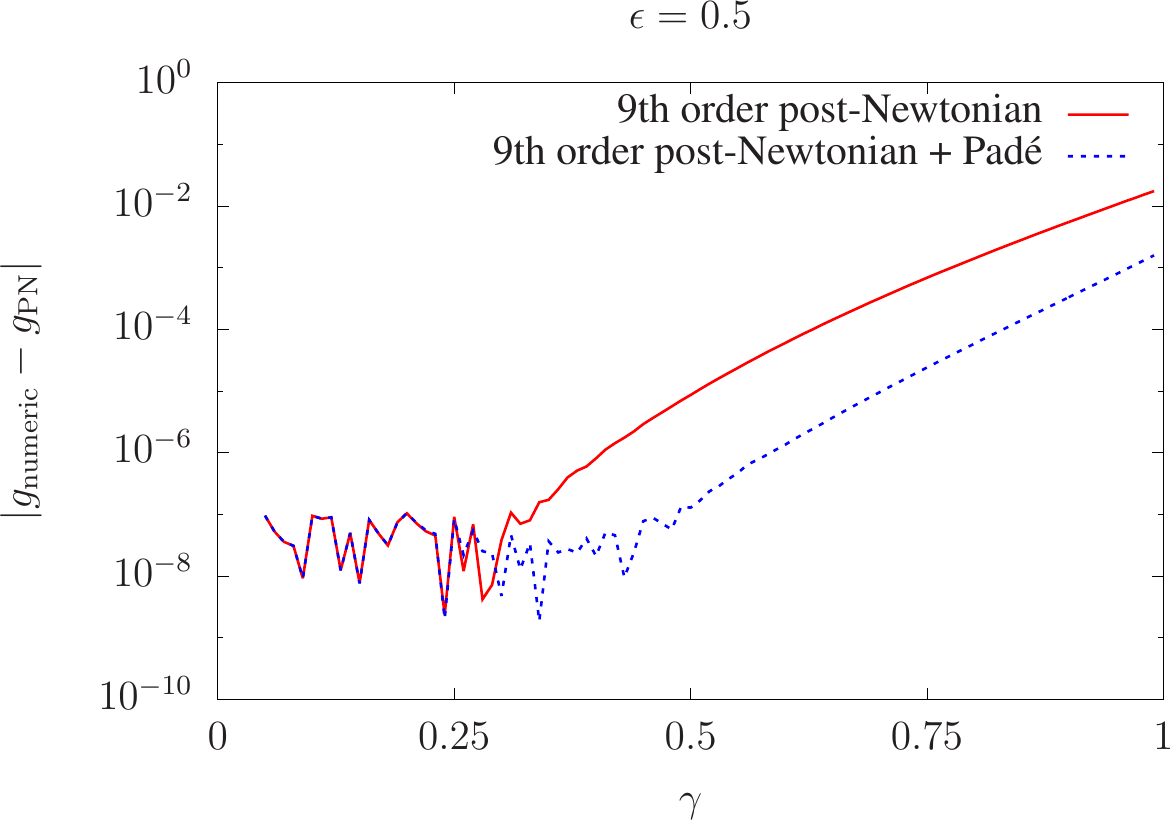}
\end{center}
\caption{Left: Comparison of the numerical and post-Newtonian result. The upper left window depicts the detail of the $g$-factor near the Newtonian limit, whereas the lower right shows the result near the ultrarelativistic limit. Right: The comparison of the accuracy between the numerical and the post-Newtonian/Pad\'{e} approximation.}
\label{fig:gyromagnetic_PN_Comparison}
\end{figure*}
 
Figure \ref{fig:gyro_versus_gamma} brings a further inset, where we zoom in on the ultra-relativistic limit $\gamma \approx 1$. Note that the $g$-factor monotonically approaches the black-hole limit and the value $g = 2$ is achieved with slope zero, i.e.,  
\[ \lim_{\gamma \rightarrow 1} g_{,\gamma} = 0\]

First hints for this behavior can be alluded from a high-order post-Newtonian expansion~\cite{Breithaupt:2015xva}. Yet, the ultrarelativistic limit is rather delicate. Hence, an ultimate conclusion regarding this issue requires the use of more powerful techniques. Indeed, we compared our numerical results with the one obtained via a post-Newtonian expansion up to the 9th order\footnote{In appendix \ref{App:gfactor_PN} we present concretely the expansion of the $g-$factor up to the 4th order.}. Here we show results for the case $\epsilon = 0.5$.  

The left panel of fig.~\ref{fig:gyromagnetic_PN_Comparison} shows that, as expected, the post-Newtonian method reproduces the behavior at low $\gamma$, in particular the evidence of $g=1$ in the Newtonian limit $\gamma \ll 1$. However, the post-Newtonian expansion is not so accurate in the prediction of the $g$-factor at larger $\gamma$ (see lower inset). A more detailed comparison is depicted in the right panel of fig.~\ref{fig:gyromagnetic_PN_Comparison}, where we display the difference between the numerical $g$-factor $g_{\rm numeric}$ and the post-Newtonian $g_{\rm PN}$.

Note that in~\cite{Breithaupt:2015xva}, the extrapolation for larger $\gamma$ values is addressed with two techniques. One either directly calculates the series expansion in the parameter $\gamma$ or one makes use of a Pad\'{e} extrapolation with the obtained coefficients. Both methods are displayed in fig.~\ref{fig:gyromagnetic_PN_Comparison}. Confirming our previous explanation, the numerical results deliver a more accurate description as we increase $\gamma$. In the post-Newtonian best performance (post-Newtonian expansion together with Pad\'{e} extrapolation), the numerical solution is more accurate from $\gamma \approx 0.5$ onwards and the error in the black-hole limit is of the order $10^{-3}$.

\section{Discussion}\label{sec:Discussion}
In this work we calculated the gyromagnetic ratio of rotating disks of electrically charged dust. The disk is parametrized by a specific charge $\epsilon$ and a parameter $\gamma$ controlling the strength of relativistic effects. The system is modelled with an energy-momentum tensor composed of dust and an electromagnetic contributions and the resulting Einstein-Maxwell equations are solved numerically with spectral methods.

This system provides us with a nice scenario to study and discuss the $g-$factor of electrically charged rotating objects in both Newtonian and relativistic regimes. Indeed, our highly accurate numerical results showed that the $g-$factor approaches the classical value $g = 1$ in the Newtonian limit $\gamma \ll 1$ while the black-hole value $g=2$ is obtained in the ultrarelativistic limit $\gamma \rightarrow 1$. In particular, these two values are connected smoothly and monotonically through the parameter $\gamma$. The dependence on $\epsilon$, on the other hand, is rather mild.

While in this work we focused on a rigidly rotating disk, we would like to stress that our approach imposes no restriction to this feature and one could use the same setup (field equations and boundary conditions) to obtain numerical solutions with differential rotation $\Omega = \Omega(\rho)$. In a broader perspective, it would be interesting to address the question under which general conditions the relation $1\le g \le 2$ holds. Studies in these directions are planned for future work.

\section*{Acknowledgments}
The authors would like to thank Marcus Ansorg, Michael Kalisch, Andreas Kleinw\"achter and Gernot Neugebauer for valuable discussions, and Christopher J Pynn for proofreading the manuscript. This work was supported by the Deutsche Forschungsgemeinschaft (DFG--GRK 1523/2). R.P.M. was supported by CNPq under the programme ``Ci\^{e}ncia sem Fronteiras". Y.-C.P. is additionally supported by the Ministry of Education, Taiwan under the program ``Government Scholarship to Study Abroad (GSSA)''.

\begin{appendix}
 \section{Field equations in $\{\sigma,\tau\}-$coordinates}\label{app:Eqs_SpecCoord}
 In this first appendix we display the equations numerically implemented in terms of the spectral coordinates $\{ \sigma, \tau\}$ [see eq.~\eqref{eq:SpecCoord}]. For the numerical solution, it is convenient to introduce the re-scaled fields $$\tilde{\omega}=\dfrac{\rho_0}{\nu^{\rm c}}\omega, \quad {\rm and} \quad \tilde{A}_{t}=\dfrac{\rho_0}{\nu^{\rm c}}A_t.$$
 In the electro-vacuum region $(\sigma,\tau)\in (0,1)^2$, the field equations \eqref{eq:EOM1}-\eqref{eq:EOM4} read
 \begin{widetext}
 \bea
 \label{eq:EOM1_newcoord} 
&&{\rm Eq}_{\nu}:-\frac{1}{2}{\rho_0}^2 (1- \tau) {\nu}F^{\rm Dust}_{{\nu}}+F^{\rm EM}_{{\nu}}=0,
   \label{E_M_eq_1_sigma_tau} \qquad {\rm with}  \\
&&\qquad F^{\rm Dust}_{{\nu}}=(1-3\tau){\nu}_{,\tau}+2\Delta'{\nu}-\frac{2}{{\nu}}\left(\nabla'{\nu}\right)^2-\frac{(1-\tau)(\nu^{\rm c})^2}{\cos^2\left(\frac{\pi}{2}\sigma\right){\nu}^3}\left(\nabla'\tilde{\omega}\right)^2,\notag\\
&&\qquad F^{\rm EM}_{{\nu}}=\left[\cos^2\left(\frac{\pi}{2}\sigma\right){{\nu}}^4+(\nu^{\rm c})^2(1-\tau){\tilde{\omega}}^2\right]\left(\nabla'A_{\varphi}\right)^2+(\nu^{\rm c})^2(1-\tau)\left[2\tilde{\omega}\nabla' A_{\varphi}\nabla'\tilde{A_t}+\left(\nabla'\tilde{A_t}\right)^2\right] \nn \\
\nn \\
\label{eq:EOM2_newcoord}
&&{\rm Eq}_{\tilde{\omega}}: -\frac{{\rho_0}^2 (1 - \tau)}{8 \cos ^2\left(\frac{\pi }{2}\sigma\right){{\nu}}^2}F^{\rm Dust}_{\tilde{\omega}}+F^{\rm EM}_{\tilde{\omega}}=0, \qquad {\rm with} \nn \\
&&\qquad F^{\rm Dust}_{\tilde{\omega}}=\frac{1}{\pi}\sin(\pi\sigma)\tilde{\omega}_{,\sigma}+(1-5\tau)\tilde{\omega}_{,\tau}-\frac{8}{{\nu}}\nabla'{\nu}\nabla'\tilde{\omega}+2\Delta'\tilde{\omega},\notag\\
 &&\qquad F^{\rm EM}_{\tilde{\omega}}=\left[\nabla'A_{\varphi}\nabla'\tilde{A_t}+\tilde{\omega}\left(\nabla'A_{\varphi}\right)^2\right],  \\
\nn \\
&&{\rm Eq}_{\tilde{A}_t}:
 \label{eq:EOM3_newcoord} 
\frac{1}{2}(1-3\tau){\nu}\left(\tilde{\omega}A_{\varphi,\tau}+\tilde{A_t}_{,\tau}\right)-2\tilde{\omega}\nabla'{\nu}\nabla'A_{\varphi}+{\nu}\nabla'\tilde{\omega}\nabla'A_{\varphi}\notag\\
 &&\qquad-2 \nabla'{\nu}\nabla'\tilde{A_t}+{\nu}\tilde{\omega}\Delta'A_{\varphi}+{\nu}\Delta'\tilde{A_t}=0.
\\
\nn \\
\label{eq:EOM4_newcoord}
&&{\rm Eq}_{A_\phi}: \frac{1}{2\pi}\cos^2\left(\frac{\pi}{2}\sigma\right){\nu}^4\big[\sin(\pi\sigma)A_{\varphi,\sigma}-\pi(1-\tau)A_{\varphi,\tau}\big] \notag\\
 &&\qquad -2\cos^2\left(\frac{\pi}{2}\sigma\right){\nu}^3\nabla'{\nu}\nabla'A_{\varphi}+(\nu^{\rm c})^2(1-\tau)\tilde{\omega}\nabla'\tilde{\omega}\nabla'A_{\varphi}\notag\\
 &&\qquad +(\nu^{\rm c})^2\left[(1-\tau)\nabla'\tilde{\omega}\nabla'\tilde{A_t}\right]-\cos^2\left(\frac{\pi}{2}\sigma\right){\nu}^4\Delta'A_{\varphi}=0.
 \eea
\end{widetext}
The action of the operators $\Delta'$ and $\nabla'$ onto two generic function $a(\sigma, \tau)$ and $b(\sigma, \tau)$ is, respectively,
\bea
 &\nabla'a\nabla'b:=\dfrac{1}{\pi^2}\cos^2\left(\dfrac{\pi}{2}\sigma\right)a_{,\sigma}b_{,\sigma}+\tau(1-\tau)a_{,\tau}b_{,\tau}, \nn \\
 &\Delta'a:=\dfrac{1}{\pi^2}\cos^2\left(\dfrac{\pi}{2}\sigma\right)a_{,\sigma\sigma}+\tau(1-\tau)a_{,\tau\tau}. \nn
\eea
  
 Moreover, the equivalent to the boundary conditions \eqref{eq:BoundA}-\eqref{eq:BoundC},\eqref{eq:BoundD2} and \eqref{eq:BoundD3} are, respectvely 
\bit
\item Region ${\cal A}:  \tau = 1, \sigma\in (0,1)$ 
\bea
\label{eq:BC_A_newcoord}
\lim\limits_{\tau\rightarrow 1} \dfrac{{\rm Eq}_{\nu}}{1-\tau}, \,
\lim\limits_{\tau\rightarrow 1} \dfrac{{\rm Eq}_{\tilde{\omega}}}{1-\tau}, \, 
\left.{\rm Eq}_{\tilde{A}_t}\right|_{\tau = 1}, \,
A_\phi= 0. 
\eea
\item Region ${\cal B}:  \sigma = 1, \tau \in [0,1]$
\bea
\label{eq:BC_B_newcoord}
\nu = 1, \,
\omega = A_t = A_\phi = 0.
\eea
\item Region ${\cal C}: \tau = 0, \sigma\in[0,1)$
\bea
\label{eq:BC_C_newcoord}
\begin{array}{c}
\nu_{,\sigma} = \tilde\omega_{,\sigma} = A_\phi{}_{,\sigma} = \tilde{A}_\phi{}_{,\sigma} = 0 \quad (\sigma = 0) \\
\\
\left. {\rm Eq}_{\nu}\right|_{\tau = 0},  \left. {\rm Eq}_{\tilde\omega}\right|_{\tau = 0}, \left. {\rm Eq}_{A_{\phi}}\right|_{\tau = 0}, \left. {\rm Eq}_{\tilde{A}_t}\right|_{\tau = 0} \quad {\rm (else)}
\end{array}
\eea
\item Region ${\cal D}:  \sigma = 0, \tau \in (0,1]$:
\bea
&& 4\nu\tilde{V}{\nu}_{,\sigma} = - \nu^{\rm c}\left[ 1 + V^2\right]\,\tilde{\omega}_{,\sigma} , \nn \\
\label{eq:BC_D_newcoord}
&& 4\nu^3A_\phi{}_{,\sigma} = \epsilon \rho_0 \nu^{\rm c}(1-\tau)\sqrt{1 -V^2}\,\tilde{\omega}_{,\sigma}, \\
&& 4\nu\tilde{V}\left[ \tilde{A}_t{}_{,\sigma} + \tilde\omega A_{\phi}{}_{,\sigma}  \right]= - \epsilon \rho_0 \sqrt{1-V^2}\,\tilde{\omega}_{,\sigma}, \nn \\
&& \nu (1+V^2) \nu_{,\tau} = - \dfrac{1}{2}\left[\nu^2 \tilde{V} + 2\nu^{\rm c}(1-\tau) \tilde{\omega}_{,\tau} \right]\tilde{V} \nn \\
&& +  \nu \frac{\epsilon}{\rho_0}\, \left[ A_{\phi}{}_{,\tau} \left(\nu^{\rm c} \tilde\omega + \tilde{V}\nu^2 \right) + \nu^{\rm c} \tilde{A}_t{}_{,\tau}\right]\,\sqrt{1-V^2}. \nn
\eea
\eit
Note that we introduced
$$
\tilde{V} = \dfrac{V}{\sqrt{1-\tau}} = \dfrac{\Omega \rho_0 - \nu^{\rm c}\tilde{\omega}}{\nu^2}
$$
in the expressions above. 

In order to complete the system, we must also fix $\nu^{\rm c}$ at the center of the disk. This value is related to the relativity parameter $\gamma$ via \eqref{eq:gamma}. Thus, at $( \sigma = 0, \tau =1 )$, we impose an extra condition
\beq
\label{eq:ExtraCond}
\nu(0,1) = \nu^{\rm c} = 1 - \gamma.
\eeq

 \section{Post-Newtonian expansion for the $g$-factor}\label{App:gfactor_PN}
 With the help of studies in the post-Newtonian approximation from~\cite{Meinel:2012wm,palenta_meinel_2013,Breithaupt:2015xva}, we expand the $g-$factor in the form\footnote{We encountered a misprint in eq.~(A.23) from~\cite{Breithaupt:2015xva}. One of the terms proportional to $\psi^4$ should read $9/22400$ instead of $9/86400$.}
\beq
 g=1 + \sum_{k=0}^{\infty}\sum_{\ell=0}^{k} c_{k,\ell} \, \epsilon^{2\ell}\, \gamma^k. \\
\eeq
 Here, we present the coefficients up to 4th order in $\gamma$:
 \begin{widetext}
 \begin{align}
&c_{1,0}=\frac{38}{35} , \quad 
 c_{1,1}=\frac{1}{5}, \quad 
 c_{2,0}=\frac{80}{9 \pi ^2}-\frac{1181}{1575},\quad 
 c_{2,1}=\frac{20789}{66150}-\frac{40}{9 \pi ^2}, \quad 
 c_{2,2}=-\frac{19}{675}, \notag \\
 \notag \\
&c_{3,0}=-\frac{19277}{808500}-\frac{592}{3465 \pi ^2}, \quad
 c_{3,1}=\frac{8891257}{16632000}-\frac{2936}{495 \pi ^2}, \quad
 c_{3,2}=\frac{2152}{693 \pi ^2}-\frac{539977931}{1629936000}, \quad
 c_{3,3}=\frac{260177}{12936000}-\frac{8}{45 \pi ^2}\notag \\
  \notag \\
&c_{4,0}=-\frac{7978729279}{22702680000}-\frac{56320}{243 \pi ^4}+\frac{22923716}{868725 \pi ^2}, \quad
  c_{4,1}=\frac{369388881091199}{488198430720000}+\frac{26240}{81 \pi ^4}-\frac{41255480963}{1021620600 \pi ^2}\notag \\
    \notag \\
&c_{4,2}=-\frac{579496867964537}{976396861440000}-\frac{29120}{243 \pi ^4}+\frac{36234236351}{2043241200 \pi ^2},\quad
c_{4,3}=\frac{1320650497820209}{10252167045120000}+\frac{640}{81 \pi ^4}-\frac{3807395827}{2043241200 \pi ^2}\notag \\
\notag \\
&c_{4,4}=\frac{104}{1215 \pi ^2}-\frac{5474341391}{715134420000}. \notag
 \end{align}
 \end{widetext}
As already mentioned, we clearly recover the classical result $g=1$ as $\gamma=0$. Moreover, we also obtain a finite limit in both uncharged $(\epsilon = 0)$ and electrically counterpoised case $(\epsilon = 1)$. 
\end{appendix}

\bibliography{GyroMagneticChargedDisk}
\end{document}